\documentclass[article,twocolumn,
amsmath,amssymb,aps,prl
]{revtex4}

\usepackage{color}
\usepackage{graphicx}
\usepackage{dcolumn}
\usepackage{braket}
\usepackage{bm}
\usepackage{amsmath,amssymb,amsthm}
\usepackage{lipsum} 
\usepackage{hyperref}
\usepackage{comment}
\usepackage{braket}

\begin{document}

\author{Filippo Caleca}
\affiliation{%
ENS de Lyon, Universit\'e Lyon 1, CNRS, Laboratoire de Physique, F-69342 Lyon, France
}%
\author{Saverio Bocini}
\affiliation{%
ENS de Lyon, Universit\'e Lyon 1, CNRS, Laboratoire de Physique, F-69342 Lyon, France
}%
\author{Fabio Mezzacapo}
\affiliation{%
ENS de Lyon, Universit\'e Lyon 1, CNRS, Laboratoire de Physique, F-69342 Lyon, France
}%
\author{Tommaso Roscilde}
\affiliation{%
ENS de Lyon, Universit\'e Lyon 1, CNRS, Laboratoire de Physique, F-69342 Lyon, France
}%

\title{Giant number-parity effect leading to spontaneous symmetry breaking in finite-size quantum spin models}

\begin{abstract}
Spontaneous symmetry breaking (SSB) occurs when a many-body system governed by a symmetric Hamiltonian, and prepared in a symmetry-broken state by the application of a field coupling to its order parameter $O$, retains a finite $O$ value even after the field is switched off. SSB is generally thought to occur only in the thermodynamic limit $N\to \infty$ (for $N$ degrees of freedom). In this limit, the time to restore the symmetry once the field is turned off, either via thermal or quantum fluctuations, is expected to diverge. Here we show that SSB can also be observed in \emph{finite-size} quantum spin systems, provided that three conditions are met: 1) the ground state of the system has long-range correlations; 2) the Hamiltonian conserves the (spin) parity of the order parameter;  and 3) $N$ is odd. Using a combination of analytical arguments and numerical results (based on time-dependent variational Monte Carlo and rotor+spin-wave theory), we show that SSB on finite-size systems can be achieved via a quasi-adiabatic preparation of the ground state -- which, in U(1)-symmetric systems, is shown to require a symmetry breaking field vanishing over time scales $\tau \sim O(N)$. In these systems, the symmetry-broken state exhibits spin squeezing with Heisenberg scaling. 
\end{abstract}

\maketitle

\emph{Introduction.} It is well known that our universe is less symmetric than the physical laws that govern it, fundamentally due to spontaneous symmetry breaking (SSB) \cite{Palmer1982, Ortiz-book,Beekman_2019} -- i.e. the tendency of many-body systems to avoid the restoration of their symmetry after an external perturbation has broken it.  
The mechanism behind symmetry breaking has been the subject of many fundamental studies in the past.  This is especially relevant in the quantum realm \cite{Anderson1952,Anderson-book,Tasaki_2018,Beekman_2019}, since the only truly stationary states of Hamiltonian dynamics are the Hamiltonian eigenstates, generically possessing all the symmetries of the Hamiltonian itself. 
The seminal work of P. W. Anderson \cite{Anderson1952,Anderson-book} has identified the emergence of symmetry breaking in quantum systems as resulting from the quasi-degeneracy of low-energy states in the presence of a large number $N$ of degrees of freedom. A perturbation mixing quasi-degenerate states with the true ground state can lead to a symmetry-broken state; once the perturbation is removed, the symmetry will be restored over a time scale which is related to the inverse of the gap $\delta$ separating the low-lying states from the ground state. If this gap scales to zero for large $N$, then the time scale for symmetry restoration will diverge in the same limit. This is the case for quantum spin systems possessing a continuous (e.g. $U(1)$ or $SU(2)$) symmetry, and exhibiting a so-called Anderson tower of states in the spectrum with $\delta \sim O(1/N)$ \cite{Anderson1952,Tasaki_2018,Beekman_2019}; and even more so for systems with a discrete (e.g. $Z_2$) symmetry, for which $\delta \sim O(\exp(-N))$ \cite{Damski_2014}. 

\begin{figure}[ht!]
	\centering
	\includegraphics[width=0.9\columnwidth]{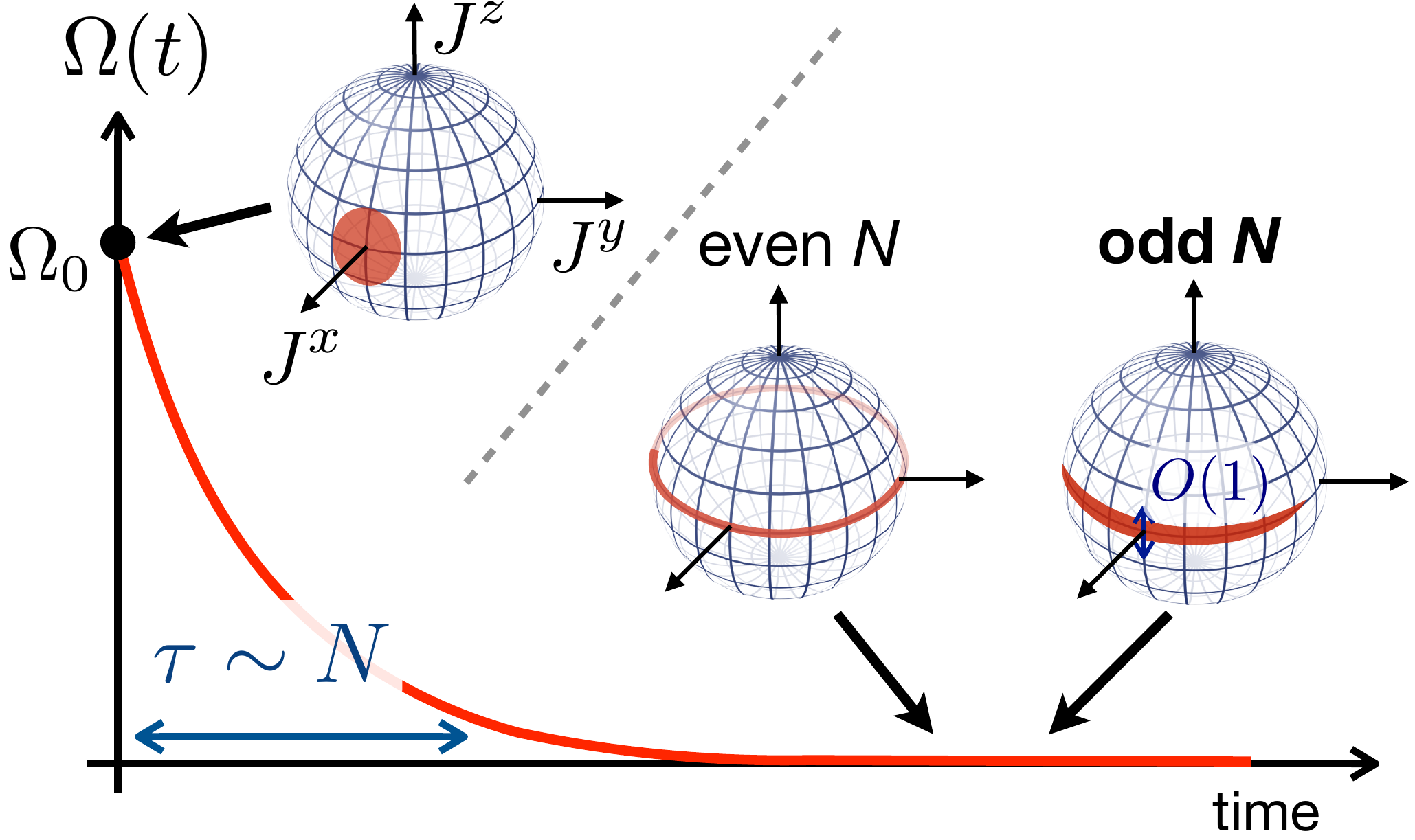}
	\caption{\emph{Spontaneous symmetry breaking from a giant number-parity effect.} Ramping down a field $\Omega$ coupling to the order parameter, a system of $N$ quantum spins can be driven (quasi-)adiabatically from a coherent spin state to a state which retains a macroscopic order parameter if: $N$ is odd; the Hamiltonian conserves spin parity; and it stabilizes long-range order in its ground state. Here we sketch the case of a $U(1)$-symmetric Hamiltonian, leading to spin squeezing in the final state: while here we picture the state on the collective-spin Bloch sphere, for the models we consider in this work the state lives actually inside the sphere, as the collective spin is not necessarily of maximal length.}
	\label{fig:sketch}
\end{figure}

Recent experiments on synthetic quantum matter -- such as ultracold atoms \cite{Gross2017}, Rydberg-atom arrays \cite{BrowaeysL2020}, trapped ions \cite{Monroe2021RMP}, superconducting circuits \cite{Juanjobook} -- open a new perspective on symmetry breaking in many-body quantum systems, as they allow one to explore the onset of collective behavior by gradually increasing their number $N$ of degrees of freedom. In particular, the dynamical restoration of symmetry due to the finite level spacings in a finite-size system -- specifically the $O(1/N)$ finite-size gap in the Anderson tower of states -- can be directly explored \cite{Bohnet2016,Bornet_2023}.  In this work, we show that choosing an \emph{odd} number $N$ of qubits in quantum spin lattices allows one to prevent the dynamical restoration of the symmetry by inducing a perfect degeneracy of the low-lying levels. Such a degeneracy in odd-$N$ lattices is protected by spin-parity conservation, leading to the fundamental consequence that symmetries can remain broken even in finite-size systems.  We illustrate this giant number-parity effect in 2d quantum spin systems with U(1) symmetry, prepared via (quasi-)adiabatic ramps of a field coupled to the order parameter -- see Fig.~\ref{fig:sketch} for a sketch. The U(1) symmetry of the interactions, broken by the quasi-adiabatic state, leads to spin squeezing with Heisenberg scaling, achieved in a time scaling linearly with system size.

\emph{Number-parity effect under spin-parity symmetry.} In this work we consider a broad class of quantum spin Hamiltonians, namely the XYZ model with power-law interactions
\begin{equation}
H = - {\cal J} \sum_{i<j} \frac{S_i^x S_j^x + \Delta_y S_i^y S_j^y + \Delta_z S_i^z S_j^z }{r_{ij}^\alpha}  - \Omega \sum_i \epsilon_i S_i^x~.
\label{e.Ham}
\end{equation} 
Here $S_i^\mu$ ($\mu=x,y,z$) are spin-1/2 operators attached to the sites $i$ ($=1,..., N$) of a lattice with periodic boundary conditions. The lattice geometry and the sign of the interaction $\cal J$ and of the anisotropies $\Delta_{y,z}$ are generically assumed to be such that 1)  the system develops long-range order in its ground state for the $S_i^x$ spin components, namely that $\langle S_i^x S_j^x \rangle \to c \neq 0$ up to the maximum distance  $r_{ij}$ between the sites, with $c$ independent of $N$;  and 2) the ground state of the Hamiltonian in the Ising limit $\Delta_y = \Delta_z = \Omega = 0$ is only doubly degenerate. This excludes the case of certain frustrated $S_i^x S^x_j$ interactions, which may lead to a higher degeneracy.  The $\Omega$ field is then chosen to couple to the order parameter $J^x = \sum_i \epsilon_i S_i^x$. Without loss of generality, in the following we choose ${\cal J}>0$ and $|\Delta_{y,z}| \leq 1$, namely dominant ferromagnetic interactions; and hence a uniform order parameter, $\epsilon_i = 1~ \forall i$, forming together with $J^{y,z} = \sum_i S_i^{y,z}$ the components of the (uniform) collective spin operator ${\bm J}$. 

A fundamental property of the Hamiltonian of Eq.~\eqref{e.Ham} is that it conserves the spin parity of the order parameter, namely it commutes with the parity operator $P^x = \prod_{i=1}^N (2S_i^x)$. This aspect allows us to prove our first rigorous result: if $N$ is odd, the ground state of the Hamiltonian Eq.~\eqref{e.Ham} with $\Omega=0$, and satisfying the above-stated assumptions, is doubly degenerate. The proof is rather elementary and it is based on the observation that, in the limit $|\Delta_{y,z}| \ll 1$, the ground state can be built perturbatively from the two states $\ket{\Rightarrow} = \otimes_i \ket{\rightarrow}_i$ and $ \ket{\Leftarrow}= \otimes_i \ket{\leftarrow}_i$ minimizing the ferromagnetic interactions among the $S^x$ spin components. Given that the two unperturbed ground states have opposite parity for $N$ odd ($P^x \ket{\Rightarrow}=1$ and $P^x \ket{\Leftarrow}=-1$), the off-diagonal Hamiltonian terms cannot connect them, and therefore the perturbed ground states remain degenerate (and of opposite parity) to any order in perturbation theory, namely in the exact spectrum throughout the long-range ordered phase. 

In the special case of U(1) symmetry, $\Delta_y=1$ (or even SU(2) symmetry, i.e. $\Delta_z=1$), the Hamiltonian commutes with the $J^z$ operator, which can only take half-integer values for $N$ odd. Given that the Hamiltonian has inversion symmetry along $z$, the energy is an even function of the $J^z$ quantum number, and therefore the ground state has even degeneracy, since its energy is independent of the sign of $J^z$.
Moreover, the energy of the low-lying levels is a quadratic function of $J^z$, $E_{J^z} \approx (J^z)^2/(2I)$, forming the spectrum of the Anderson tower of states \cite{Anderson1952,Tasaki_2018,Beekman_2019}, with $I \sim O(N)$ (see Supplemental Material (SM) for further details \cite{SM}). When $N$ is odd, the degenerate ground states have therefore $J^z = \pm 1/2$, and we shall indicate them as $|\Psi_{\pm1/2}\rangle$. The two eigenstates of $J^z$ do not have a definite parity $P^x$, since $P^x$ and $J^z$ do not commute (in fact they anti-commute). Yet they can be rotated to two $P^x$ eigenstates, $|\pm\rangle = \left ( |\Psi_{1/2}\rangle \pm |\Psi_{-1/2}\rangle \right )/\sqrt{2}$, with $P^x |\pm\rangle = \pm |\pm\rangle$. 

We can then prove our second rigorous result: the order parameter takes a macroscopic value $\langle J^x \rangle \sim O(N)$ on the ground states of $H$ (Eq.~\eqref{e.Ham}) which are also eigenstates of $P^x$. The proof, albeit simple, is too lengthy to be offered here, and it is presented in the SM \cite{SM}. 
From the proof one can deduce that, in the U(1)-symmetric case of $\Delta_y = 1$,  $\langle J^x \rangle = \pm \frac{1}{2} \langle \sqrt{{\bm J}^2 + 1/4} \rangle$ on the $|\pm\rangle$ states; and that, if $\langle {\bm J}^2 \rangle \lesssim  N/2(N/2+1)$, such as in the limit $\alpha \to 0$ in Eq.~\eqref{e.Ham}, then $\langle J^x \rangle \approx N/4$.

\begin{figure*}[ht!]
	\centering
	\includegraphics[width=0.95\textwidth]{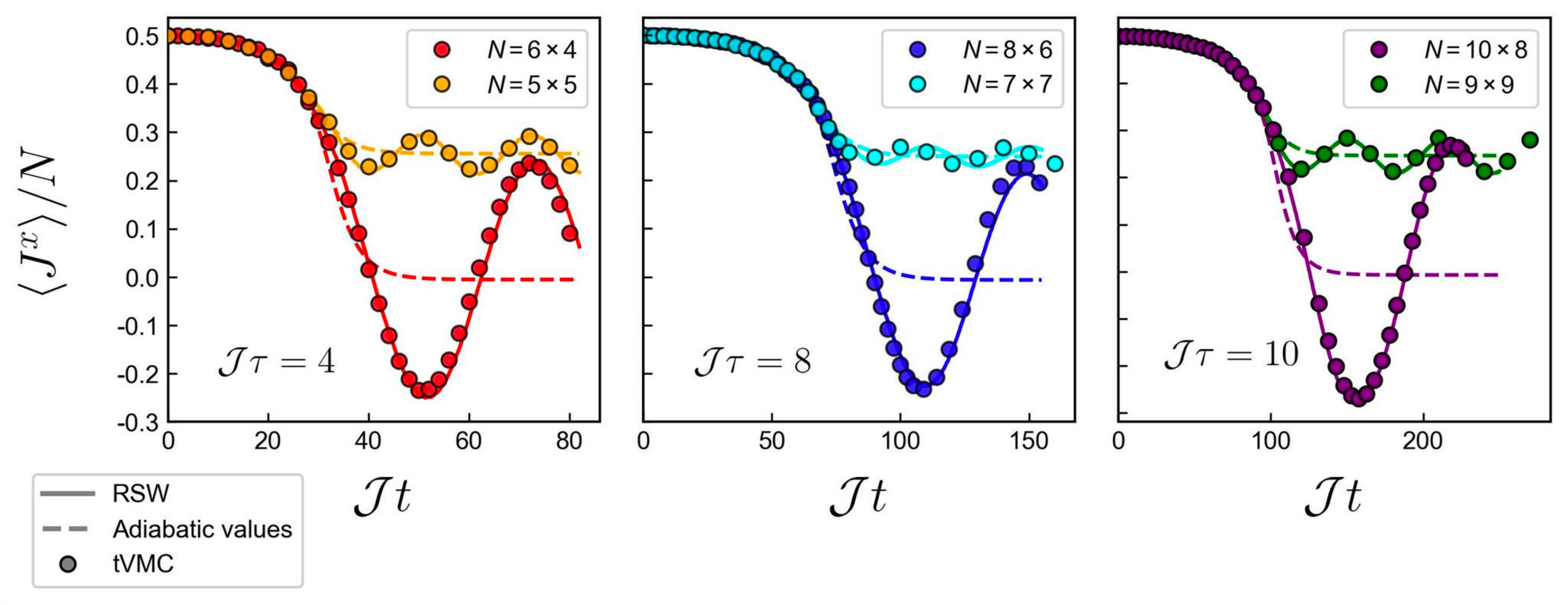}
	\caption{\emph{Spontaneous symmetry breaking at finite size in the $2d$ dipolar $XX$ model.} Evolution of the magnetization under an exponential field ramp $\Omega_0 \exp(-t/\tau)$ in the 2d dipolar XX model, for  three pairs of system sizes differing by one spin, and ramp durations $\tau {\cal J} = 4, 8$ and 10 (and $\Omega_0 = 20 {\cal J}$). Circles represent tVMC results; solid and dashed lines correspond to RSW results for the time evolution and the instantaneous ground-state (or adiabatic) value, respectively.}
	\label{fig:2d_ramps}
\end{figure*}

\emph{Finite-size spontaneous symmetry breaking from odd number parity.} 
The above results lead to the fundamental conclusion that the degenerate and  long-range ordered ground states of Hamiltonian $H$ of Eq.~\eqref{e.Ham}, cast on odd-$N$ lattice, display a macroscopic value of the order parameter when they have a well-defined spin parity.  This has important consequences for the preparation of such states, which can therefore exhibit spontaneous symmetry breaking on finite-size systems. The preparation of the state $|+\rangle$ can indeed simply proceed via a (quasi-)adiabatic ramp of the applied field $\Omega = \Omega(t)$, from $\Omega(0) \gg \cal J$ to $\Omega=0$ for times $t \gg \tau$, where $\tau$ is a characteristic ramp time scale (see Fig.~\ref{fig:sketch}). Throughout the rest of the paper, we shall choose an exponential ramp $\Omega(t) = \Omega_0 \exp(-t/\tau)$. If the system is prepared \emph{e.g.} in the coherent spin state $|\Rightarrow\rangle$ at time $t=0$, close to the ground state of the Hamiltonian Eq.~\eqref{e.Ham} with $\Omega \gg \cal J$, then the subsequent evolution will conserve the (positive) parity of the state, and, if sufficiently slow, it will prepare a state close to $|+\rangle$, displaying a macroscopic order parameter $\langle J^x \rangle$. This protocol clearly realizes spontaneous symmetry breaking on a finite-size lattice, namely the persistence of a finite order parameter in the stationary state of a system governed by a symmetric Hamiltonian. It is important to stress that this property of persistence of $\langle J^x \rangle \sim O(N)$ is \emph{not} a finite-size transient, but it is a truly stationary condition, since $|+\rangle$ is a Hamiltonian eigenstate. 

The condition on the scaling of $\tau$ with $N$ which is necessary to ensure adiabaticity will be discussed at length later. Here we show instead the manifestation of spontaneous symmetry breaking in odd-$N$ lattices in an actual quasi-adiabatic evolution, contrasting it with what happens in the case of an even $N$. To this end, we consider a rectangular $L_1\times L_2$ lattice of dipolar spins ($\alpha=3$) with XX interactions, namely $\Delta_y = 1$ and $\Delta_z=0$ in Eq.~\eqref{e.Ham}. This case is directly relevant to dipolar Rydberg-atom arrays, for which quasi-adiabatic preparations of low-energy states were recently reported, albeit with a different protocol that cannot fix the spin parity of the resulting state \cite{Chen_2023}. 

\begin{figure*}[ht!]
	\centering
	\includegraphics[width=\textwidth]{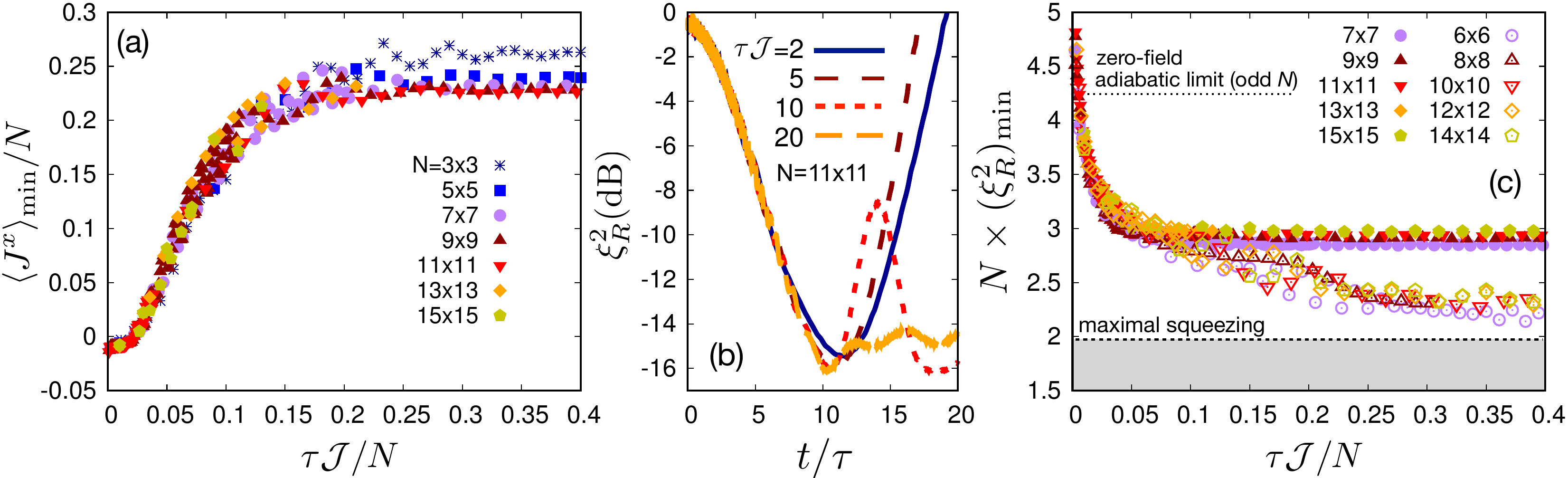}
	\caption{\emph{Dynamical scaling for SSB and spin squeezing.} (a) Minimum residual magnetization per spin during an exponential ramp with $\Omega_0 = 20 \cal J$ as a function of the ramp time $\tau$. The actual ramp duration is $t\gg \tau$; (b) Evolution of the squeezing parameter $\xi_R^2$ for a $N=11 \times 11$ lattice and for various ramp durations; (c) Minimum squeezing parameter $\xi_R^2$ during the same ramps; the dashed line marks the minimum value for $\xi_R^2$ allowed by quantum states. All data were obtained for the 2d dipolar XX model using RSW theory.}
	\label{fig:tau_scaling}
\end{figure*}

We calculate the evolution of the quantum state of the system under an exponential ramp of the $\Omega$ field by making use of two independent approaches: 
time-dependent Variational Monte-Carlo (tVMC) based on the pair-product (or spin-Jastrow) wavefunction  \cite{PhysRevB.100.155148,Comparin_2022} supplemented with a magnetization-dependent term (see SM \cite{SM});
and rotor+spin-wave (RSW) theory \cite{Roscildeetal2023}, extending spin-wave theory to include the full nonlinear dynamics of the zero-mode excitations building up the Anderson tower of states. Both approaches are very accurate in the study of the quench dynamics of the 2d dipolar XX model \cite{Comparin_2022_cats,Roscildeetal2023};  here we apply them to dynamics at much lower energies, but for significantly longer times -- ideally reaching adiabatic time scales. 
Fig.~\ref{fig:2d_ramps} shows the results of our calculations, exhibiting a very good agreement between the two techniques we use, and fully displaying the giant number-parity effect. Indeed rectangular lattices differing by a single site exhibit completely different dynamics:
 
\noindent I) The odd-sized lattices show SSB, with a residual magnetization which is very close to the value $\approx N/4$ for infinite-range interactions (see above); and which oscillates around the adiabatic value. The amplitude of the oscillations is controlled by the residual excitation energy at the end of the ramp, and it is suppressed when both $\Omega_0$ and $\tau$ increase. For the ramps we used, the state at times $t \gg \tau$ can be written as $|\Psi(t)\rangle \approx \alpha e^{-iE_{1/2} t} |+\rangle + \beta e^{-iE_{3/2} t} |+_{3/2}\rangle$ with
$|+_{3/2}\rangle = \left (|\Psi_{3/2}\rangle + |\Psi_{-3/2}\rangle \right)/\sqrt{2}$, namely it admixes the ground state with the first excited states in the Anderson tower (at $J^z = \pm 3/2$). For this state, $\langle J^x \rangle(t) = |\alpha|^2 \langle J_x \rangle_+ + |\beta|^2 \langle J_x \rangle_{+,3/2} + |\alpha| |\beta| \langle \Psi_{3/2} | J^+ |\Psi_{1/2} \rangle \cos\left (\frac{t}{I}-\phi_{\alpha\beta} \right )$, with $\langle... \rangle_{+(,3/2)} = \langle +_{(3/2)} | ... | +_{(3/2)} \rangle$ and $\phi_{\alpha\beta}$ the phase difference between the $\alpha$ and $\beta$ coefficients;

\noindent  II) On the other hand, the even-sized lattices show a magnetization that exhibits large fluctuations around zero -- its adiabatic value in the case of a unique ground state  $|\Psi_{J^z=0}\rangle$ with $J^z = 0$. The amplitude of the fluctuations is again controlled by the admixture of the ground state with the first two states of the Anderson tower, namely $|\Psi_{J^z=\pm 1}\rangle$. For $t \gg \tau$ with the ramps we used, the state reads $|\Psi(t)\rangle \approx \alpha e^{-iE_0 t} |\Psi_0\rangle + \beta e^{-iE_1 t} |+_1\rangle$ with $|+_1\rangle = \left (|\Psi_1\rangle + |\Psi_{-1}\rangle \right)/\sqrt{2}$, for which  $\langle J^x \rangle(t) = \sqrt{2} |\alpha| |\beta| \langle \Psi_{1} | J^+ |\Psi_{0} \rangle \cos\left (\frac{t}{2I}-\phi_{\alpha\beta} \right )$, namely the frequency of the oscillations is nearly half of that for the odd-sized case, since $I_{N-1} = I_{N}(1- O(1/N))$.  The oscillation amplitude should only be $\approx \sqrt{2}$ larger than in the odd-size case, $\alpha$ and $\beta$ being the same, while it appears to be much larger in Fig.~\ref{fig:2d_ramps}. This must be related to the fact that the excitation probability $|\beta|^2$ is larger in the even-sized case, given that the excitation gap $E_1 - E_0  = \frac{1}{2I}$ is nearly half of that in the odd-sized case, $E_{3/2}-E_{1/2} = \frac{1}{I}$.  

These results are rather general for systems exhibiting long-range order in the ground state, as long as the field ramps lead to moderate excitations in the final state -- in the SM \cite{SM} we show similar results for the case of the 2d XX model with nearest-neighbor interactions ($\alpha=\infty$). What remains to be determined is the condition on $\tau$ to achieve the quasi-adiabatic dynamics leading to SSB on odd-sized lattices.  

\emph{Time scale to adiabaticity.} In order to test the conditions on the ramp timescale $\tau$ to achieve SSB, we repeat our study of finite-size lattices for a variable $\tau$, testing which $\tau$ values lead to a macroscopic residual magnetization at the end of the ramp. For $\tau \to 0$, the ramp dynamics is equivalent to a quantum quench, which in general does not admit any residual magnetization -- e.g. in the limit of $\alpha=0$ with $\Delta_y=1$ and $|\Delta_z|<1$ the quench dynamics generated by $H$ of Eq.~\eqref{e.Ham} is the exactly solvable one-axis-twisting dynamics, for which the magnetization evolves as $\langle J^x \rangle = (N/2)\cos^{N-1}[t/(2I)]$ \cite{Ma2011PR}, converging exponentially to zero for sufficiently long times ($t\gtrsim \sqrt{N}$) regardless of the parity of $N$. This implies that a minimal value for $\tau$ is required for the ramp dynamics to sustain a residual magnetization and hence SSB for odd $N$ values.  We perform a systematic study of ramps with different time scales $\tau$ and for a range of system sizes (up to $N=15\times 15$) by making use of RSW theory -- which is very reliable for the dynamics of interest in light of its agreement with the tVMC results, as seen in Fig.~\eqref{fig:2d_ramps}. Fig.~\ref{fig:tau_scaling}(a) shows the minimum value $\langle J^x \rangle_{\rm min}/N$ attained by the order parameter for $t\in [0,t_0]$ with $t_0\gg \tau$ (typically $t_0 = 25\tau$), in the case of the dipolar 2d XX model . We observe that the data of the residual magnetization per spin obtained for different ramp durations and sizes collapse when rescaling the ramp duration by the system size. This implies that the quasi-adiabatic preparation of states with a given residual extensive magnetization is attained when $\tau$ scales linearly with $N$. This is a much slower scaling than the prediction $\tau \sim \delta_{\rm min}^{-2} \sim N^2$ from a na\"ive application of the adiabatic theorem \cite{Albash2018RMP}, where $\delta_{\rm min} \sim N^{-1}$ is the minimum gap during the ramp, related to the tower-of-state spectrum (see SM \cite{SM} for further discussion). In particular, the threshold value for the onset of SSB is estimated at around $\tau {\cal J} \approx 0.025 N$, while the maximum value for the residual magnetization is attained for $\tau {\cal J} \gtrsim 0.2 N$. The maximum value falls short of the value expected from a perfect adiabatic evolution, simply because of the finite initial field $\Omega_0$ ($=20 {\cal J}$ in Fig.~\ref{fig:tau_scaling}) which implies that a residual excitation energy is present from the very beginning of the evolution. 

\emph{Heisenberg scaling of spin squeezing.} If the Hamiltonian has U(1) symmetry, as in the example discussed above, the adiabatic evolution of odd-sized lattices brings the system to the $|+\rangle$ state, which, besides having a macroscopic magnetization $\langle J^x \rangle_+ \sim O(N)$, also displays the remarkable property that the fluctuations of the $J^z$ collective-spin component are microscopic, since $J^z$ can only take values $\pm 1/2$ -- hence ${\rm Var}_+(J^z) = 1/4$. As a consequence, the spin squeezing parameter \cite{Wineland1994PRA,Pezze2018RMP}
\begin{equation}
\xi_R^2 = \frac{N \min_\theta {\rm Var}(J^\theta)}{\langle J^x \rangle^2}
\end{equation}
with $J^\theta = \cos\theta J^z + \sin\theta J^y$, is of order $O(1/N)$ in the $|+\rangle$ state -- for which $\min_\theta {\rm Var}(J^\theta) = {\rm Var}(J^z)$. Spin squeezing, namely the condition $\xi_R^2 < 1$, is a multipartite entanglement property, with $\xi_R^2 < 1/k$ (for $k$ an integer) implying $(k+1)$-partite entanglement \cite{Pezze2018RMP}. Hence $k\sim N$ gives the fastest possible scaling (the so-called Heisenberg scaling) for the spin squeezing parameter. More specifically, we have seen that $\langle J^x \rangle_+\leq (N+1)/4$, so that $(\xi_R^2)_+ \geq 4N/(N+1)^2$, which misses only by a factor $\approx 2$ the minimum possible value $2/(N+2)$ authorized for quantum states \cite{Pezze2018RMP}. 
Hence finite-size SSB due to the giant parity effect is accompanied by spin squeezing with Heisenberg scaling, when the dynamics prepares adiabatically the $|+\rangle$ state. Yet this property is robust to an imperfect (i.e. quasi-adiabatic) preparation of the state exhibiting SSB, as shown in Fig.~\ref{fig:tau_scaling}(c). There we observe that the optimal value of the squeezing parameter $(\xi_R^2)_{\rm min}$, corresponding to the first global minimum of  $\xi_R^2(t)$ along an exponential ramp of the applied field (see Fig.~\ref{fig:tau_scaling}(b)), obeys Heisenberg scaling if the ramp time  $\tau$ grows linearly with $N$; and that for $\tau {\cal J} \gtrsim 0.1 N$, the fastest scaling (in terms of the prefactor to the $1/N$ decay) is already attained. And this scaling is actually faster than the $\gtrsim 4/N$ scaling (for $N\gg1$) attained with the adiabatic preparation \cite{SM}. This result witnesses the fact that $\xi_R^2(t)$ has an intermediate minimum which is lower than the value at long times, even for the longest ramps (see Fig.~\ref{fig:tau_scaling}(b); and see SM \cite{SM} for the adiabatic behavior). 

\emph{Robustness to particle loss.} The notion of a giant parity effect might erroneously suggest that a fine control on the particle number $N$ is needed to observe the physics we have discussed in this work -- and this would be very detrimental, since in any quantum simulation platform (e.g. in neutral-atom arrays) quasi-adiabatic preparations can come with particle losses/ qubit decays. In fact, it is clear that losing particles simply leads to a reduction of the residual magnetization at the end of the field ramp -- which would be averaged between the value for odd $N$, oscillating around a finite value; and the value for an even $N$, oscillating around zero. The Heisenberg scaling of spin squeezing obtained with the quasi-adiabatic ramps is potentially robust to particle losses, since the same scaling is featured by the optimal squeezing in the case of even $N$, as shown in Fig.~\ref{fig:tau_scaling}(c) (see also SM \cite{SM}). Hence the observation of the effects discussed in this work does not require the post-selection of experiments based on the final number of particles.  

\emph{Conclusions.} In this work we have unveiled a giant number-parity effect leading to spontaneous symmetry breaking (SSB) in finite-size systems, namely the persistence of a macroscopic magnetization in odd-sized quantum spin lattices after a field stabilizing the magnetization is slowly turned off. This effect requires spin Hamiltonians which conserve parity; and which possess long-range order in the ground state. As such it can be observed in a broad variety of quantum simulation platforms (ultracold atoms in optical lattices, Rydberg-atom arrays, trapped ions, superconducting circuits, etc.) generically realizing the XYZ model on lattices and/or with interaction ranges that can stabilize long-range order. The extension of this effect to Luttinger liquids in one-dimensional systems will be discussed in a forthcoming publication. A further interesting extension would be to ensembles of spins with $S>1/2$. Our results extend the picture of the scalable adiabatic preparation of spin-squeezed states in U(1) symmetric systems, put forward in Ref.~\cite{Comparinetal2022}, by showing that Heisenberg scaling of spin squeezing can in fact be achieved within a time scaling linearly with system size, and in fact for shorter time scales than those required to achieve strict adiabaticity. Hence our work paves the way for a possible metrological exploitation of low-energy states in systems exhibiting the SSB phenomenon \cite{Comparinetal2022,Blocketal2023}.

\acknowledgements{\emph{Acknowledgements.} We acknowledge fruitful discussions with G. Bornet, A. Browaeys, C. Chen, G. Emperauger and T. Lahaye, whose experimental results have sparked this study. This work is supported by PEPR-q ("QubitAF" project). All numerical simulations were run on the CBPsmn cluster of the ENS of Lyon.}

\bibliography{bibliography_SSB}

\begin{thebibliography}{28}
\bibitem[{SM()}]{SM}
\bibinfo{note}{See {S}upplemental {M}aterial ({SM}) for details about: 1) the
  tower-of-states spectrum of even and odd $N$; 2) the proof that parity
  eigenstates with odd $N$ have a macroscopic magnetization; 3) the
  spin-Jastrow Ansatz used in the tVMC calculations; 4) the magnetization
  dynamics in the 2d XX model with nearest-neighbor interactions; 5) the
  adiabatic values of the squeezing parameter; 6) the condition on adiabaticity
  of the exponential ramps; and 7) the absence of finite-size SSB in the
  quantum Ising model.}
\end{thebibliography}

\newpage
\null
\newpage

\section*{SUPPLEMENTAL MATERIAL}
\noindent{\bf \emph{Giant number-parity effect leading to spontaneous symmetry breaking in finite-size quantum spin models}}

\section{Tower-of-state spectrum for even and odd $N$}
\begin{figure}[ht!]
	\centering
	\includegraphics[width=\columnwidth]{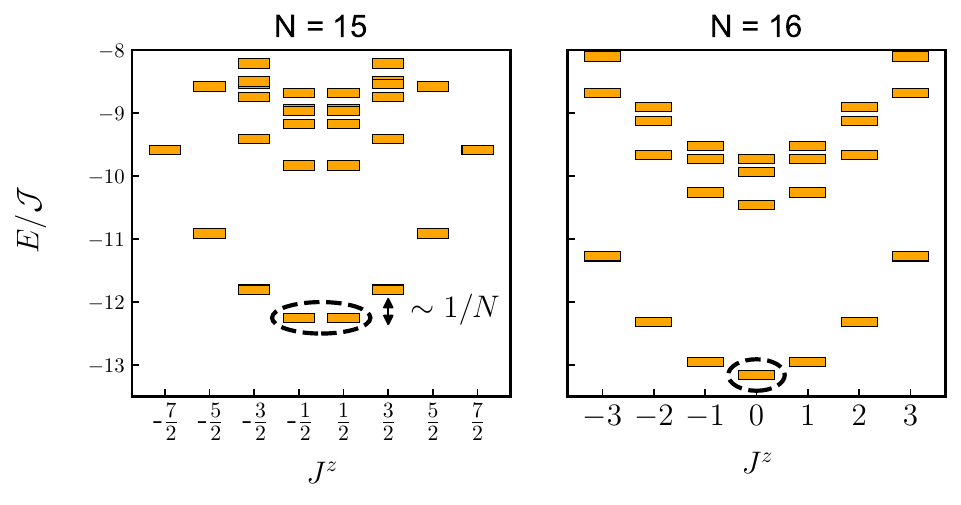}
	\caption{\emph{Tower-of-state spectrum.}  Low-lying spectrum of the dipolar XX model on a $5\times 3$ square lattice (left panel) and on a $4\times 4$ square lattice (right panel). The eigenenergies are plotted as a function of the magnetization $J^z$.}
	\label{fig:ToS}
\end{figure}

Fig.~\ref{fig:ToS} shows the exact low-lying spectrum of the dipolar XX model on two small square lattices, with an odd number of sites ($N = 15$) and with an even number of them ($N=16$). As discussed in the main text, the ground state of the odd-sized lattice is doubly degenerate, corresponding to the sectors with minimal magnetization, $J^z = \pm 1/2$; while the even-sized lattice has a unique ground state with $J^z=0$. The gap separating the ground state to the first excited state in the Anderson tower of states -- made of the ground states in the various $J^z$ sectors -- is twice as large in the odd-sized case, $\Delta E = 1/I$, as it is in the even-sized case, $\Delta E = 1/(2I)$, with $I \sim O(N)$ the moment of inertia.

 \begin{figure*}[ht!]
	\centering
	\includegraphics[width=\textwidth]{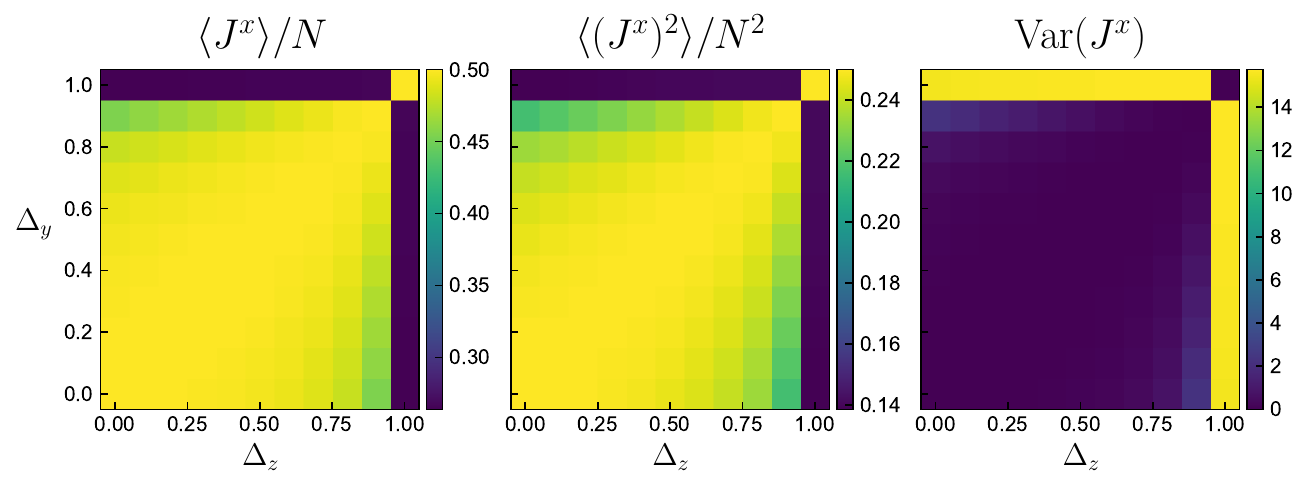}
	\caption{\emph{Order parameter and its fluctuations in the 2d dipolar XYZ model.} (a) Average order parameter; (b) average of the squared order parameter; (c) order-parameter fluctuations. All calculations refer to the exact ground state of a $3\times 5$ square lattice with positive $J^x$ parity. As discussed in the text, for every value of the $\Delta_z$ anisotropy, the U(1) symmetric case $\Delta_y = 1$ features the lowest value of the order parameter and its strongest fluctuations. }
	\label{fig:XYZ}
\end{figure*}

\section{Proof that $\langle J^x \rangle \sim O(N)$ for parity eigenstates with odd $N$ and long-range order}

Here we prove that the ground states $|\pm\rangle$ of the Hamiltonian in Eq.~1 of the main text with zero field ($\Omega=0$), which are also eigenstates of the parity $P^x$ (with eigenvalue $\pm 1$), have a macroscopic magnetization $\langle J^x \rangle \sim O(N)$ under the assumption of long-range order, implying that $\langle S_i^x S_j^x \rangle \to c \neq 0$ for $|i-j|\to \infty$. 

Without loss of generality, we shall prove that $\langle J^x \rangle_+ \sim O(N)$ where $ \langle ... \rangle_+ = \langle + | ... | + \rangle $. The central element of the proof is given by the decomposition  
of the $|+\rangle$ states on the joint eigenstates of the ${\bm J}^2$ and $J^z$ operators, $|J,M,\lambda\rangle$, for which ${\bm J}^2 \to J(J+1)$, $J^z \to M$ and $\lambda$ is a set of additional quantum numbers.  The decomposition reads generically
\begin{equation}
|+\rangle = \sum_{J=1/2}^{N/2} \sum_{J^z=1/2}^J \sum_\lambda c_{J, J^z ,\lambda} \left ( |J,J^z,\lambda\rangle +  |J,-J^z,\lambda\rangle \right)
\end{equation}
where $c_{J, J^z ,\lambda}$ can be taken as real numbers; and where we have made explicit the parity of the state by symmetrizing the superposition over the inversion of $J^z$, namely the fact that 
$c_{J, J^z ,\lambda} = c_{J, -J^z ,\lambda}$. 

In particular, in the Hamiltonian Eq.~1 of the main text, we assume that the $S^z$ spin components do not develop long-range ferromagnetic correlations, namely $\langle J^z \rangle = 0$ and ${\rm Var}(J^z) \lesssim O(N)$. This does not exclude the possibility of having long-range antiferromagnetic correlations for them -- as in a SU(2)-symmetric XXZ model -- but our main concern is with the fluctuations of the uniform collective spin component $J^z$. Hence we can safely assume that the probability distribution for the $J^z$ variable,  $P(J^z) = \sum_{\lambda, J} | c_{J, J^z ,\lambda} |^2$, is peaked around the origin, namely around $J^z = \pm 1/2$, with a width of at most $O(\sqrt{N})$. This means that  $P(J^z=\pm 1/2) \gtrsim O(N^{-1/2})$. This is what happens in the most general case of the XYZ model. In the case of the XXZ model with U(1) or even SU(2) symmetry, such that $[H,J^z]=0$, $J^z = \pm 1/2$ are the only two possible values with $P(J^z=\pm 1/2)= 1/2$, so that ${\rm Var}(J^z) = 1/4$. 

Based on the above decomposition, we can then calculate 
 \begin{align}
& \langle J^x \rangle_+ =   \sum_J p_{J,1/2} \sqrt{J(J+1)+1/4}  \\
& + \sum_J \sum_{J^z=3/2}^{J} \sum_\lambda c_{J,J^z,\lambda} c_{J,J^z-1,\lambda} \sqrt{J(J+1) - J^z(J^z-1)} \nonumber 
\label{e.Jx+}
\end{align}
and 
\begin{align}
& \langle (J^x)^2 \rangle_+ =  \sum_J p_{J,1/2} [J(J+1)-1/4]  \\
& + \sum_J \sum_{J^z=3/2}^{J } p_{J,J^z} \left [ J(J+1) - (J^z)^2 \right ] \nonumber \\
& + \frac{1}{2} \sum_J \sum_{J^z=3/2}^{J} \sum_\lambda  c_{J,J^z,\lambda} c_{J,J^z-2,\lambda} \nonumber \\  
& \sqrt{J(J+1)-J^z(J^z-1)} \sqrt{J(J+1)-J^z(J^z-2)}~.  \nonumber
\label{e.Jx2+}
\end{align}
where we have introduced the notation $p_{J,J^z} = \sum_\lambda |c_{J,J^z,\lambda}|^2$. 

In the case of U(1) or SU(2) symmetry, for which $p_{J,1/2} = p_J/2$ where $p_J = \sum_{J^z} p_{J,J^z}$, the above expressions simplify to
 \begin{eqnarray}
\langle J^x \rangle_+ & = &   \frac{1}{2} \sum_J p_{J} \sqrt{J(J+1)+1/4}  \\
\langle (J^x)^2 \rangle_+ & = & \sum_J p_{J} \left [ J(J+1) - 1/4 \right ] \nonumber
\label{e.U1}
\end{eqnarray}

 \begin{figure*}[ht!]
	\centering
	\includegraphics[width=0.9\textwidth]{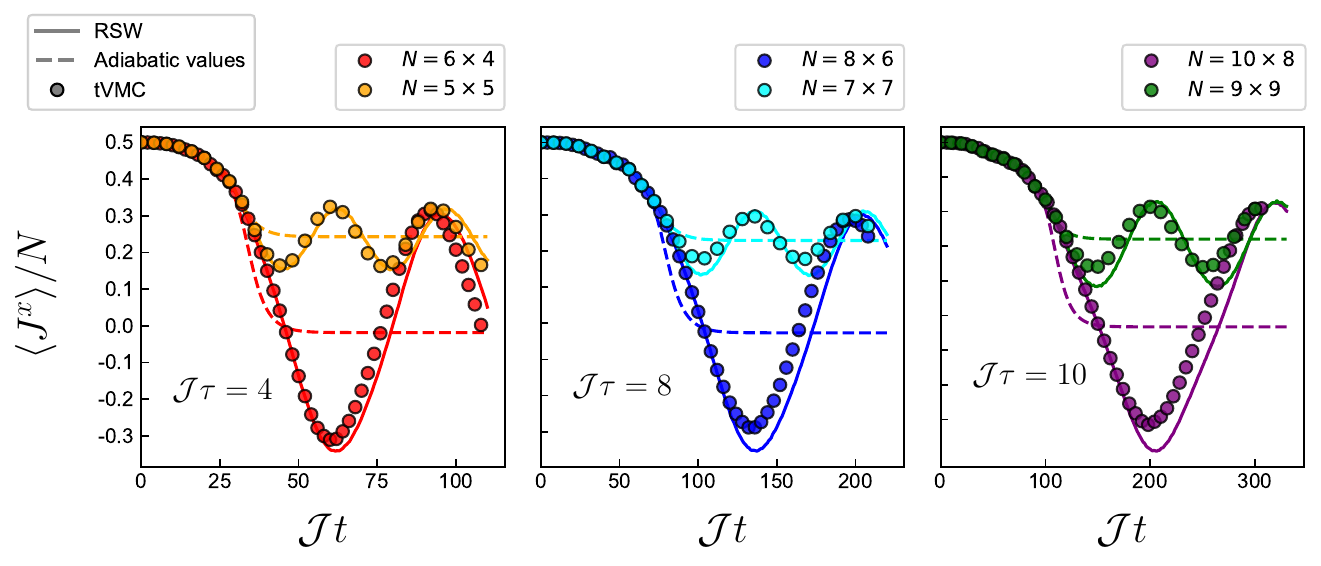}
	\caption{\emph{Spontaneous symmetry breaking at finite size in the 2d XX model with nearest-neighbor interactions.}  Magnetization dynamics along exponential field ramps with decay rate $\tau$ for three pairs of system sizes, differing by one single spin. The odd sizes show a persistent magnetization even after the field stabilizing it has been switched off. The same symbols as in Fig.~2 of the main text are used.}
	\label{fig:nnXY}
\end{figure*}

The hypothesis of long-range order implies that, to leading order,  $\langle (J^x)^2 \rangle_+ \approx a N^2$ for some $a>0$. This property implies in turn that 
\begin{align}
& a N^2 \approx \langle (J^x)^2 \rangle_+ = \sum_J p_J  J^2  (1+1/J-1/(2J)^2) \nonumber \\
& \leq  3 \sum_J p_J  J^2 
\end{align}
where we have used the fact that $J \geq 1/2$. As a consequence
\begin{equation}
\sum_J p_J  J^2 \geq a' N^2
\end{equation}
where $a' \approx a/3$. Now from Eq.~\eqref{e.U1} we deduce that 
\begin{equation}
\langle J^x \rangle_+ \geq \frac{1}{2} \sum_J p_J J 
\end{equation}
and moreover, since $J \leq N/2$ 
\begin{equation}
\sum_J p_J  J^2 \leq \frac{N}{2} \sum_J p_J J
\end{equation}
from which we obtain that 
\begin{equation}
\langle J^x \rangle_+ \geq \frac{1}{2} \sum_J p_J J \geq a' N
\end{equation}
which proves that $\langle J^x \rangle_+ \sim O(N)$. 

Hence we have come to the proof for the case of U(1) (or even SU(2)) symmetry. The same proof for the case of a $Z_2$ symmetry -- namely for the XYZ model -- appears to be more involved. Indeed in the expression of $\langle J^x \rangle_+$, one needs to include the further coherence terms in the second line of Eq.~\eqref{e.Jx+}, whose sign is not known a priori.  

Yet we can extend the proof to the case of $Z_2$ symmetry for the XYZ model by using a physical argument. Let us consider a given target XYZ model with $\Delta_y < 1$, and $|\Delta_z| \neq 1$, and let us consider a reference XXZ model with the same $\Delta_z$ but with $\Delta_y=1$, for which the above proof applies. Then moving from the XXZ model to the XYZ model upon lowering $\Delta_y$, we can follow adiabatically the positive-parity ground state, remaining in the same phase with long-range order for the $S^x$ spin component. Lowering the $\Delta_y$ terms leads necessarily to a decrease of the fluctuations of $J^x$ compared to the U(1)-symmetric case, namely to a smaller ${\rm Var}(J^x)$. On the other hand, the interactions among the $S^x$ spin components become even more dominant, leading to an increase of the $\langle S^x S^x\rangle$ correlations and hence of $\langle (J^x)^2 \rangle$. As a result, we have that the XYZ model with $\Delta_y < 1$ is expected to exhibit a larger  $\langle (J^x)^2 \rangle$ but a smaller ${\rm Var}(J^x)$ with respect to the $\Delta_y=1$ limit -- see Fig.~\ref{fig:XYZ} for a numerical evidence of this statement. This implies that $\langle J^x \rangle = \sqrt{\langle (J^x)^2 \rangle - {\rm Var}(J^x)}$ must have increased with respect to the XXZ limit, and it must be a fortiori $\sim O(N)$.

\section{Spin-Jastrow state with a magnetization-dependent term}

In this work, we have carried out our tVMC simulations using a generalization of the spin-Jastrow (or pair-product) Ansatz \cite{PhysRevB.100.155148, Comparin_2022, Comparin_2022_cats} which includes a magnetization-dependent term, namely an Ansatz of the form $|\Psi\rangle = \sum_{\bm \sigma} \psi(\bm \sigma) |\bm \sigma \rangle$ with
\begin{equation}
 \psi(\bm \sigma)  =: {\cal N} \exp\Big (\sum_{i<j} u_{ij} \sigma_i \sigma_j \Big ) ~C_M~. 
\end{equation}
Here $|\bm \sigma \rangle$ is a joint eigenstate of the $S_i^z$ operators; $u_{ij}$ are $O(N)$ coefficients in a translationally invariant system; $C_M$, with $M= {\sum_i \sigma_i}$, are $N+1$ (resp. $N$)  multiplicative coefficients for $N$ even (resp. $N$ odd) depending on the uniform magnetization along $z$ of the system; while ${\cal N}$ is an overall {normalization coefficient}. In practice, the $C_M$ coefficients reduce only to $N/2+1$ (resp. $N/2$) when the system has inversion symmetry along $z$, i.e. $C_M = C_{-M}$. The introduction of the $C_M$ coefficients is necessary to reproduce exactly {the dynamics dictated by Eq.~1 of the main text in the limit of all-to-all connected interactions (or $\alpha = 0$). The coefficients of our Ansatz are evolved in time according to the time-dependent variational principle \cite{Becca_Sorella_2017}.}


\section{Results for the 2d XX model with nearest-neighbor interactions}

Fig.~\ref{fig:nnXY} shows the tVMC and RSW results for the magnetization dynamics in the 2d XX model with nearest-neighbor interactions, during exponential ramps of the applied $\Omega$ field with decay rate $\tau$. As seen in the main text for the case of the dipolar 2d XX model, lattices differing by a single size exhibit radically different dynamics, with the odd-sized lattices showing a persistent magnetization (modulo oscillations), i.e. spontaneous symmetry breaking on a finite size. The agreement between tVMC and RSW is still rather good, although it is worse on the bigger lattices than in the case of the dipolar XX model. This is due to the increased role of spin waves, so that the assumption of their complete decoupling from the rotor degree of freedom is less accurate -- see also Ref.~\cite{Roscildeetal2024} for a study of the quench dynamics (i.e. an infinitely short ramp) in the same model.

 \begin{figure*}[ht!]
	\centering
	\includegraphics[width=0.9\textwidth]{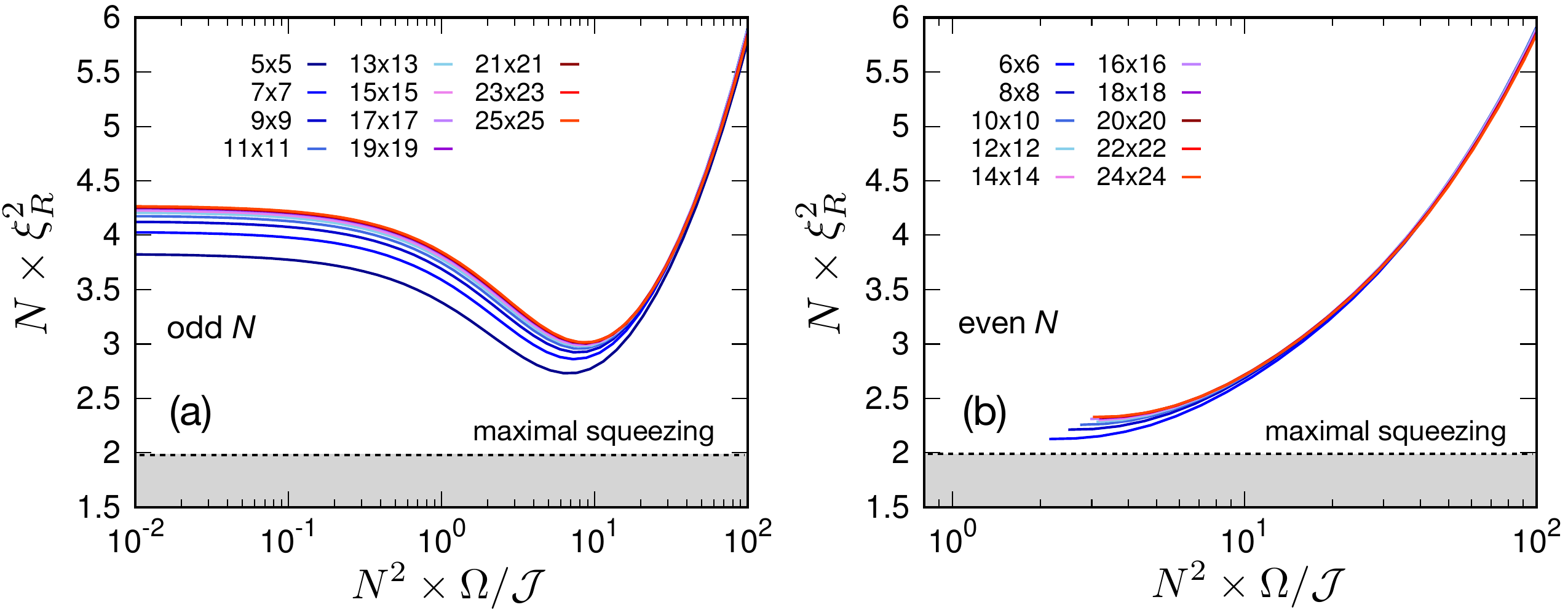}
	\caption{\emph{Adiabatic squeezing and its Heisenberg scaling.} Ground-state squeezing parameter $\xi_R^2$ for the 2d dipolar XX model in an applied field $\Omega$ from RSW theory for (a) odd $N$; and (b) even $N$.}
	\label{fig:adiabatic}
\end{figure*}

\section{Squeezing in the ground state for perfectly adiabatic ramps}

Fig.~\ref{fig:adiabatic}(a) shows the ground-state results for the squeezing parameter $\xi_R^2$, as obtained from rotor+spin-wave (RSW) theory \cite{Roscildeetal2023} applied to the 2d dipolar XX model. These static results offer an important reference for the dynamical results presented in the main text, and they show the limit of infinitely long (i.e. perfectly adiabatic) ramps. Fig.~\ref{fig:adiabatic}(a) shows that, for odd-$N$ lattices, the squeezing parameter as a function of the field exhibits an intermediate minimum, scaling approximately as $\approx 3/N$, for fields $\Omega \sim N^{-2}$. This minimum squeezing is also realized dynamically by ramps with $\tau \gtrsim 0.05 N /\cal J$ (see Fig.~3(c) of the main text). On the other hand, the zero-field limit is higher, $\xi_R^2 \approx 4.25/N$. In any instance, the Heisenberg scaling of squeezing is a robust feature of ground state squeezing, provided that one looks at fields scaling as $\Omega \sim N^{-2}$. 

Similar observations apply to the case of even-$N$ lattices, shown in Fig.~\ref{fig:adiabatic}(b), where we see a minimum squeezing scaling as $\approx 2.4/N$. This is also the scaling exhibited by the longest ramps ($\tau \gtrsim 0.25 N / {\cal J}$),  as shown in Fig.~3(c) of the main text.

 \begin{figure*}[ht!]
	\centering
	\includegraphics[width=0.95\textwidth]{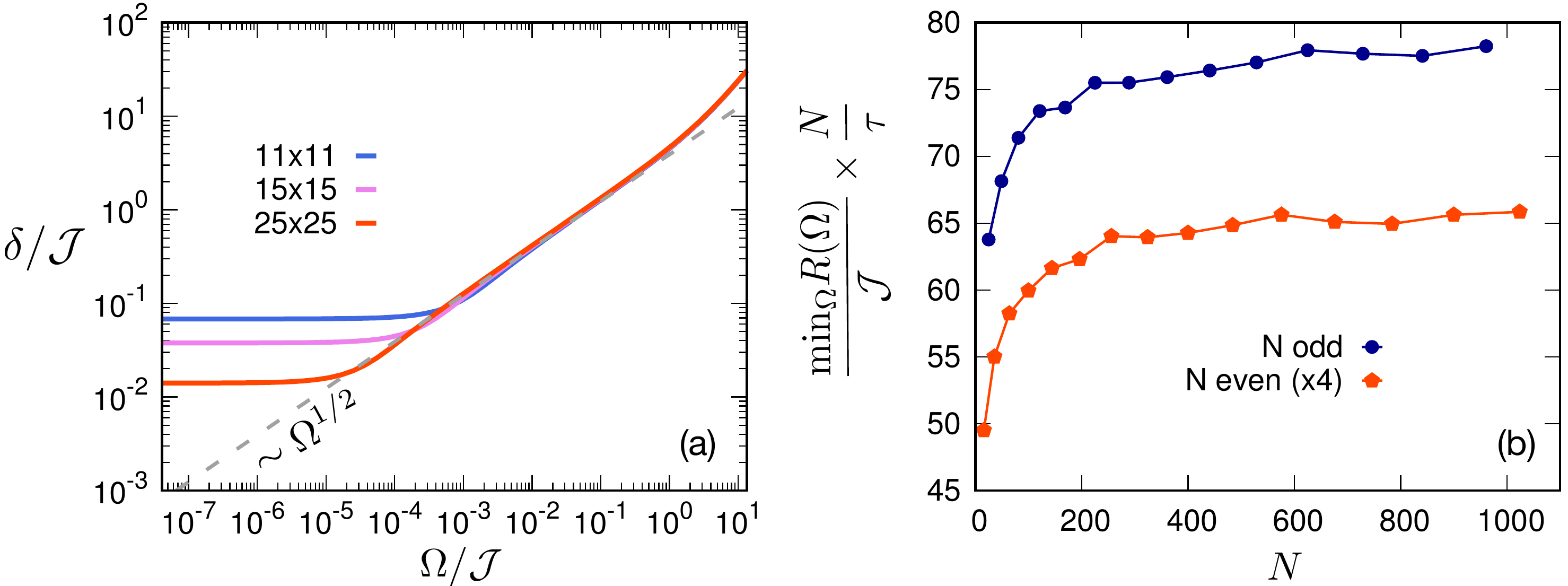}
	\caption{\emph{Conditions for adiabaticity of the field ramps.} (a) Energy gap $\delta$ over the ground state as a function of the applied field for the 2d dipolar XX model; (b) for the same model: minimum ratio $R$ (see text) for exponential ramps with decay time $\tau$, as a function of the system size. All data are obtained via RSW theory.}
	\label{fig:gap}
\end{figure*}

\section{Condition of adiabaticity for the exponential ramps}

The rigorous conditions of validity of the adiabatic theorem in quantum mechanics represent an intricate subject -- see Ref.~\cite{Albash2018RMP} for a review. 

A general, yet rather restrictive rule of thumb for adiabaticity implies that the total time of the field ramp we use should be much larger than the inverse of the square of the minimum gap 
$\delta_{\Omega(t)}$ between the ground state and the first excited state during the evolution. This gap can be estimated quantitatively using RSW theory, which predicts it to be that associated with the first excited state of the Anderson tower, namely with a rotor (or zero-momentum) excitation.
The gap is shown in Fig.~\ref{fig:gap}(a) to decrease as $\delta_\Omega \sim \sqrt{\Omega}$ upon lowering the applied field, as also predicted within ordinary spin-wave theory in the thermodynamic limit \cite{Roscildeetal2023}. The gap stops scaling with the field when $\Omega \sim N^{-2}$, namely when the field-induced gap $\sim \Omega^{1/2}$ becomes comparable with the tower-of-states gap $\sim 1/N$.

Taking the adiabatic criterion stated above at face value (as some of us did in Ref.~\cite{Comparinetal2022}) would imply that, to reach adiabatically field values of order $\Omega$, it takes ramps of characteristic duration $\tau \sim \delta^{-2}_{\Omega} \sim 1/\Omega$. As seen in the previous paragraph, in order to reach \emph{e.g} the Heisenberg scaling $\xi_R^2 \sim N^{-1}$ of the squeezing parameter in the ground state one needs to go to fields of the order $\Omega_{\rm H} \sim N^{-2}$, and hence the above rule would imply that ramps with $\tau \sim N^2$ would be needed, which is a rather prohibitive scaling. 

In fact, as shown in Fig.~3 of the main text, ramps with $\tau \sim N$ (actually, $\tau \sim 10^{-2} N / {\cal J}$) already realize the Heisenberg scaling, showing that the above criterion is far too restrictive. A less restrictive, and more precise criterion \cite{Messiah2014, Albash2018RMP} for the field ramp $\Omega(t)$ to keep the state of the system close to the instantaneous ground state within the time range  $[0,t_0]$ is that the \emph{instantaneous} matrix element between the ground state $|\psi_0\rangle$ and first excited state $|\psi_1\rangle$ of the derivative of the ramp Hamiltonian be much smaller than the square of the \emph{instantaneous} Hamiltonian gap $\delta_{\Omega(t)}$ between the same states. In mathematical terms, one defines the ratio 
\begin{equation}
R(t) = \frac{\delta_{\Omega(t)}^2}{|\dot{\Omega}(t) \langle \psi_0(\Omega(t)) | J^x |  \psi_1 (\Omega(t)) \rangle |}
\end{equation}
where  $|  \psi_{0/1} (\Omega(t)) \rangle$ are the instantaneous ground/excited states of the Hamiltonian. 
The criterion for adiabaticity then reads
\begin{equation}
\min_{t\in [0,t_0]} R(t)  \gg 1~. 
\label{e.criterion}
\end{equation}

In fact, in the case of an exponential ramp, one has that $|\dot{\Omega}(t)| = |\Omega(t)|/\tau$, so that the above condition simply reads
\begin{equation}
\min_{\Omega} \frac{\delta_{\Omega}^2}{\Omega |\langle \psi_0(\Omega) | J^x |  \psi_1 (\Omega) \rangle | \cal J } =  \min_{\Omega} \frac{R(\Omega)}{\tau \cal J}  \gg \frac{1}{\tau \cal J}~. 
\label{e.Rmin}
\end{equation}
To test whether this condition is met by our ramps, we show in  Fig.~\ref{fig:gap}(b) the value of  $ r_{\min} = \min_{\Omega} R(\Omega) /(\tau {\cal J})$  as a function of system size for both odd $N$ and even $N$, in the case of exponential ramps for the 2d XX dipolar model, as obtained from RSW theory. There we observe that $r_{\rm min} \sim 1/N$ since $ N r_{\rm min} $ is seen to scale to a constant which is $\sim 16$ for even sizes and $\sim 80$ for odd sizes.   
This observation implies that ramps with $\tau {\cal J}/N \sim O(1)$ fully satisfy the condition Eq.~\eqref{e.Rmin}, since for them  $ N r_{\rm min} \gg 1$.  In particular, $r_{\rm min}$ for the odd sizes is roughly 4 times larger than for even sizes. This observation may be related to the fact that the energy gap in the Anderson tower of states for odd sizes is $\delta = 1/I_N$, while it is $\delta = 1/(2 I_N)$ for even sizes. The minimum ratio is realized for fields $\Omega \sim N^{-2}$, such that the size dependence nearly disappears in the ratio $\delta^2/\Omega$.  

The above results fully justify the observation of the main text that ramps of duration $\tau \approx \tau_0 N$ ensure an adiabatic following of the ground state. Moreover, they also give an indication about the prefactor $\tau_0$. Indeed, this can be estimated as the inverse of $r_{\rm min} N$, namely $\tau_0 {\cal J} \gtrsim 10^{-2}$ for odd $N$; and $\tau_0 {\cal J} \gtrsim 6 \times 10^{-2}$ for even $N$.  Fig.~3 of the main text shows that in practice ramps that exceed these minimal durations by at least a factor of 5 are adiabatic. 
 Our results also invite to reconsider the conclusions that some of us reached in Ref.~\cite{Comparinetal2022}. There, using the strict criterion for adiabaticity $\tau > (\min \delta)^{-2}$, it was stated that ramps of duration $\tau \sim N$ are required to reach adiabatically fields $\Omega \sim N^{-1}$ and corresponding squeezing values $\xi_R^2 \sim N^{-1/2}$; while actually such ramp durations allow one to reach adiabatically fields  $\Omega \sim N^{-2}$ at which one achieves the Heisenberg scaling of squeezing $\xi_R^2 \sim N^{-1}$.

\section{Transverse-field Ising model: absence of finite-size spontaneous symmetry breaking} 

To put our results into context, we would like to discuss the effect of quasi-adiabatic field ramps in the case of the paradigmatic transverse-field Ising (TFI) model
\begin{equation}
H = - {\cal J} \sum_{i<j} \frac{1}{r_{ij}^\alpha} S_i^x S_j^x - h \sum_i S_i^z - \Omega \sum_i S_i^x 
\end{equation} 
where $h$ is the transverse field, and $\Omega$ is the longitudinal field coupling to the order parameter, i.e. the magnetization $J^x$. Because of the transverse field, the TFI Hamiltonian does \emph{not} conserve the parity of the order parameter, and as a result the ground state is always \emph{unique} (if $h>0$), even on odd-$N$ lattices. When $\Omega=0$ the model has a ferromagnetic phase when $h < h_c$, with a critical field $h_c$ depending on the details of the lattice and interaction geometries. In this phase the ground state $|\psi_0\rangle$  is separated from the first-excited state $|\psi_1\rangle$ by an exponentially small gap $\delta \sim O(\exp(-N))$; for $h \ll \cal J$ the two states are approximately $|\psi_0\rangle \approx ( \ket{\Rightarrow_x} + \ket{\Leftarrow_x}) / \sqrt{2}$ and $|\psi_1\rangle \approx ( \ket{\Rightarrow_x} - \ket{\Leftarrow_x}) / \sqrt{2}$. In this same phase, a slow, but not perfectly adiabatic ramp of the $\Omega$ field will bring the state of the system to a superposition of the ground and first excited state $\approx (|\psi_0\rangle + |\psi_1\rangle)/\sqrt{2}$, which exhibits a macroscopic magnetization $\langle J^x \rangle \sim O(N)$. Nonetheless, over time scales $t \sim O(\exp(-N))$ the magnetization will decay because of the relative phase accumulated by the two components of the state. This means that spontaneous symmetry breaking on a finite-size system is not realized by the TFI model, highlighting the crucial role played for this phenomenon by the conservation of the parity of the order parameter. 

\end{document}